\documentclass[twocolumn,showpacs,aps]{revtex4}

\usepackage{graphicx}
\usepackage{dcolumn}
\usepackage{bm}

\begin{document}

\title{Spatially Resolved Nonlinearity Measurements of YBa$_2$Cu$_3$O$_{7-\delta}$ Bi-crystal Grain Boundaries}
\author{Sheng-Chiang Lee}
 \email{sycamore@wam.umd.edu}
\author{Steven M. Anlage}%
 \email{anlage@squid.umd.edu}
\affiliation{
Center for Superconductivity Research, and Materials Research Science and Engineering Center, Department of Physics, University of Maryland, College Park, MD 20742-4111 USA}

\date{\today}

\begin{abstract}
We have developed a near-field microwave microscope to locally excite a superconducting film and measure second and third harmonic responses at microwave frequencies. We study the local nonlinear response of a YBa$_2$Cu$_3$O$_{7-\delta}$ thin film grown on a bi-crystal SrTiO$_3$ substrate. The location of the bi-crystal grain boundary is clearly identified by the microscope through higher harmonic response, and the spatial resolution is on the order of the magnetic loop probe size, about 500$\mu m$. The harmonic power and spatial resolution are successfully modeled with a one-dimensional extended Josephson junction simulation. From the model, the 2nd order harmonic response is dominated by Josephson vortex generation and flow. A geometry-free nonlinear scaling current density $J_{NL}\simeq 10^4\sim 10^5 A/cm^2$ is also extracted from the data, indicating that the grain boundary weak link is the dominant nonlinear source in this case.

\end{abstract}

\maketitle

The nonlinear properties of high-$T_c$ superconductors have been of great concern in microwave applications, although the microscopic origins of the nonlinear response still remain uncertain.\cite{Hein} All superconductors have an intrinsic nonlinearity associated with the nonlinear Meissner effect (NLME). Calculations of harmonic response based on BCS theory,\cite{Yip,Dahm} Ginzburg-Landau (GL) theory,\cite{Gittleman} and microwave field-induced modulation of the super/normal fluid densities near $T_c$\cite{Ciccarello1} have been proposed to describe the intrinsic nonlinearities of superconductors. Extrinsic sources of nonlinearity include grain boundaries,\cite{McDonald} edge effects, and weakly-coupled grains.\cite{Oates1} Many experiments have studied the intermodulation power,\cite{Balam,Wensheng} harmonic generation,\cite{TOI,Pestov} or the nonlinear surface impedance of superconductors\cite{Oates2,Ciccarello2} as a function of applied microwave power. However, most nonlinear experiments are done with resonant techniques, which by their nature study the averaged nonlinear response from the whole sample rather than locally. Such techniques usually have difficulty in either avoiding edge effects, which give undesired vortex entry, or in identifying the microscopic nonlinear sources. Therefore, a technique capable of locally measuring nonlinear properties of samples is necessary for understanding the physics of different nonlinear mechanisms. In addition, most existing experimental techniques focus on 3rd order nonlinearities, which can be conveniently studied by intermodulation techniques, but rarely address the 2nd order nonlinear response. Here we present a technique to locally characterize 2nd and 3rd order nonlinearities through spatially localized harmonic generation.

In prior work,\cite{Wensheng} we studied the intermodulation signal from a high-$T_c$ superconducting microwave resonator using a scanned electric field pick-up probe. Both the "global" and the "local" intermodulation power measured with the open-end coaxial probe were presented. However, the local measurements were actually a superposition of nonlinear responses that were generated locally but propagated throughout the microstrip and formed a resonant standing-wave pattern. To avoid this loss of spatial information, we have developed a non-resonant near-field microwave microscope, to non-destructively measure the local harmonic generation from un-patterned samples. This technique has the additional advantage of operating equally well through $T_c$ and into the normal state of the sample.

In our experiment (Fig.\ \ref{Schematic}), low pass filters are used to filter out higher harmonics generated by the microwave source, and guarantee that only the selected fundamental frequency is sent to the sample. A loop probe, which is made of a coaxial cable with its inner conductor forming a semi-circular loop shorted with the outer conductor, induces microwave frequency currents of a controlled geometry in the sample. The probe can be translated over the surface of the sample in the x-y plane. The harmonic signals generated in the sample couples back to the loop probe. An advantage of this probe is the localized and directional microwave current induced on the sample surface, which enables the study of direction-dependent nonlinearities. Reflected harmonics are selected by 2 high pass filters, before being amplified by $\sim$ 65 dB, and measured by a spectrum analyzer.

The sample is a 500$\AA$ thick YBa$_2$Cu$_3$O$_{7-\delta}$ (YBCO) thin film deposited by pulsed laser deposition on a bi-crystal SrTiO$_3$ substrate with a 30$^\circ$ mis-orientation. The distance between the loop probe and the sample is fixed by a 12.5$\mu m$ thick Teflon sheet placed between them.
Measurements of the temperature dependent 3rd order harmonic power ($P_{3f}$)are first performed, both above the grain boundary (GB) and far away from the grain boundary (non-GB), as shown in Fig.\ \ref{Tramp}. A strong peak in $P_{3f}(T)$ is observed at $T_c=88.9K$ (measured by ac susceptibility) at all locations on the sample. The peaks have similar magnitudes at both locations although there is a slight shift of $T_c$. Note that all measurements are taken near the middle of the film where we have verified that current-enhancement edge effects are absent.\cite{John} This peak near $T_c$ is predicted by all models of intrinsic nonlinearities of superconductors, e.g. the NLME,\cite{Yip,Dahm} GL theory,\cite{Gittleman} etc., and a power-3 dependence of the $P_{3f}$ on the input microwave power ($P_f$) is observed, in agreement with the predictions of those models. The order parameter of the fragile superconducting state is easily suppressed by the microwave current at temperatures near $T_c$, so that its nonlinear dependence on the external rf current dominates the microwave response. Also note that no feature is seen in $P_{2f}(T)$ near $T_c$, as expected for a time-reversal symmetric superconductor.

The SrTiO$_3$ substrate is a nonlinear dielectric at low temperatures,\cite{STO} and we have measured harmonic response from bare SrTiO$_3$ substrates below 80K.\cite{John} The nonlinear response is confined to narrow temperature ranges at temperatures when the substrate becomes resonant due to its temperature-dependent, high dielectric constant. However, at temperatures below 80K, a strongly temperature dependent $P_{3f}$ is observed above the YBCO bi-crystal grain boundary, while no detectable $P_{3f}$ is seen away from the grain boundary. Such clear contrast is evidence that the observed $P_{3f}$ is from the grain boundary, not the SrTiO$_3$ substrate. Power dependencies of $P_{2f}$ and $P_{3f}$ were also performed at both locations at 60K ($\ll T_c$) and 95K ($>T_c$). As shown in Fig.\ \ref{Pramp}, strongly power dependent $P_{2f}$ and $P_{3f}$ are only observed above the grain boundary, while no response is seen above the background noise level away from the bi-crystal grain boundary.

To demonstrate how the microwave microscope is able to spatially resolve a localized source of nonlinearity, a measurement of $P_{2f}$ and $P_{3f}$ along a line crossing the grain boundary is performed. As shown in Fig.\ \ref{linecut}, a clear peak in both $P_{2f}$ and $P_{3f}$ is observed above the GB, with a width of about 500$\mu m$ . The width of the observed $P_{2f}/P_{3f}$ peaks are about the size of the loop probe, which determines the spatial distribution of the surface current on the sample. This interpretation is confirmed by reproducing this peak with the extended resistively shunted Josephson junction model (ERSJ) discussed below.

It is well known that applying microwave current to a single resistively shunted Josephson junction generates harmonics at all odd integer multiples of the drive frequency.\cite{McDonald,VanDuzer} To obtain a more comprehensive understanding of weak link junctions, the ERSJ model was introduced to model long Josephson junctions, such as the YBCO bi-crystal grain boundary.\cite{McDonald,Oates1,Oates2} In this paper, we present ERSJ models to simulate a YBCO bi-crystal grain boundary as either an array of identical inductively coupled, or independent (uncoupled), Josephson junctions acting in parallel. From prior work with these junctions, we expect each to have a critical current of 8$\mu A$, and a shunt resistance of 20$\Omega$, and the currents applied to each junction vary according to the surface current distribution on the film induced by the loop probe. We sum up the nonlinear potential differences across all junctions, and extract the higher harmonics from this collective nonlinear potential difference via Fourier expansion.\cite{McDonald}
The spatial distribution of the surface current density is calculated from a simplified analytical model of an ideal circular loop in a vertical plane, with radius 270$\mu m$, coupling to a perfectly conducting horizontal plane 382.5$\mu m$ away from the center of the loop. The magnitude of the current density is determined by a much more sophisticated microwave simulation by the Ansoft High Frequency Structure Simulator software, which also produces a similar surface current distribution. 

The calculation of the uncoupled ERSJ model is performed by Mathematica, and is shown as the dashed line centered on 4mm in Fig.\ \ref{linecut}. This model predicts a narrow distribution of $P_{3f}$ of greater magnitude (almost 20dB) than is observed experimentally. It also predicts no $P_{2f}$, due to the absence of Josephson vortices in this model. A power dependence calculation from this uncoupled ERSJ model is also performed and compared with experimental results (dashed line in Fig.\ \ref{Pramp}). The comparison shows qualitative agreement with the $P_{3f}(P_f)$ data. Calculations with the inductively coupled ERSJ model, which includes Josephson vortices, performed by WRSpice, give very good descriptions in both magnitude and spatial resolution of the experimental results for both $P_{2f}$ and $P_{3f}$ (solid lines in Fig.\ \ref{linecut}). 

To further evaluate the capability of our microscope to detect intrinsic superconducting nonlinearities due to different mechanisms, we extract a geometry-free scaling current density, $J_{NL}$,\cite{Balam} from our data.
We assume that the quadratic nonlinearity in the kinetic inductance of our sample dominates the nonlinear response through

\begin{equation}
L_{KI}\equiv\frac{\mu_0}{I^2}\int_{volume}\lambda_{L}(T,J)^2 J^2 dV\equiv L_0+\Delta L I^2,
\end{equation}
and $P_{3f}=(\frac{\omega\Delta L I^{3}}{4})^2/(2Z_0)$, where $P_{3f}$ is the 3rd harmonic power in the sample, $\omega$ and $I$ are the microwave frequency and current, $Z_0$ is the characteristic impedance of the transmission line, $L_{KI}$ is the kinetic inductance of the sample, $L_0$ and $\Delta L$ are the linear and 2nd order nonlinear terms in $L_{KI}$, $\lambda_{L}(T,J)=\lambda_{L}(T,0)(1+(\frac{J}{J_{NL}(T)})^2)^{1/2}$ is the temperature and current dependent London penetration depth, and $J_{NL}$ is the nonlinearity scaling current density. Different microscopic models of nonlinearity predict different values and temperature dependences of $J_{NL}(T)$. For example, in the nonlinear Meissner effect and Ginzburg-Landau theory,$J_{NL}\sim10^8 A/cm^2$ or higher,\cite{Gittleman,Balam} except for temperatures close to $T_c$, while the $J_{NL}$ of a long 1-D Josephson junction array in series combination is expected to be about $10^6 A/cm^2$ or less.\cite{Balam}
In our case, we follow Booth's algorithm\cite{Booth} to calculate the nonlinear term in the inductance, and the expected 3rd order harmonic response, all as a function of $J_{NL}$, using the current distribution calculated from the model of the loop probe. However, it should be noted that the calculation of $J_{NL}$ is based on the assumption that the nonlinearity in the kinetic inductance of the superconductor is purely quadratic. As long as $P_{3f}$ does not scale with $P_{f}^{3}$, the calculated $J_{NL}$ will be power dependent.\cite{Balam} 

Assuming $\lambda_{L}(T=0,J=0)=1500\AA$, we obtain the $J_{NL}$ calculated near the grain boundary to be $J_{NL}^{GB}\sim 1.5\times 10^5 A/cm^2$, while the sensitivity of our setup is currently limited to $J_{NL}\leq 1.3\times 10^6 A/cm^2$.\cite{John} However, it is suggested from the model calculation that thinner films and stronger coupling between the film and the loop probe will give stronger nonlinear response from the same mechanism, and improve the sensitivity to the nonlinearities associated with larger values of $J_{NL}$.

We have created a novel near-field microwave microscope that is capable of locally measuring the nonlinear microwave behavior, thus providing an important new tool to study nonlinearities of superconductors. The microscope is able to distinguish GB and non-GB areas through spatially resolved nonlinearity measurements of higher harmonic generation. The spatial resolution is determined by the size of the loop probe, which is about 500$\mu m$ in the current setup, and can be improved by reducing the size of the loop probe, and decreasing the loop/sample separation. A quantitative understanding of the magnitude and spatial resolution of the 2nd and 3rd order harmonic response is well achieved by the coupled ERSJ model. A geometry-free scaling current density $J_{NL}$ is extracted from our measurement, and can be used to further study the physics of different nonlinearities of superconductors.  

We thank Su-Young Lee for making the YBCO thin films, and Greg Ruchti for the HFSS calculation. This work is supported by DARPA DSO contract $\#$ MDA972-00-C-0010 through a subcontract by STI, the University of Maryland/Rutgers NSF MRSEC under grant DMR-00-80008 through the Microwave Microscope SEF, and NSF IMR under grant DMR-98-02756.
\pagebreak

\pagebreak

\begin{figure}[p]
\caption{\label{Schematic}Experimental schematic. The 6.5 GHz microwave signal is low-pass filtered before being sent to the sample, and high-pass filtered before being amplified and measured by the spectrum analyzer.}
\end{figure}

\begin{figure}[p]
\caption{\label{Tramp}Temperature dependence of 3rd (a) and 2nd (b) harmonic generation on non-GB YBCO surface (circles) and YBCO grain boundary (GB) (triangles) in the temperature range 5$\sim$90 K. The fundamental frequency is 6.5 GHz and input power +8dBm.}
\end{figure}

\begin{figure}[p]
\caption{\label{Pramp}Power dependence of 2nd and 3rd harmonic generation on both non-GB YBCO surface and grain boundary at 60K ($\ll T_c$) (filled-triangles: $P_{3f}$ on GB; diamonds: $P_{2f}$ on GB; circles: $P_{3f}$ on non-GB; invered-traingles: $P_{2f}$ on non-GB). The uncoupled-ERSJ model calculation for the power dependence of the 3rd harmonic generation on the grain boundary at 60K is also plotted as a dashed line for comparison.}
\end{figure}

\begin{figure}[p]
\caption{\label{linecut}Second- ($P_{2f}$, circles) and third-harmonic ($P_{3f}$, triangles) power along a line crossing the bi-crystal grain boundary, which is located at X=4mm. Calculations of $P_{2f}$ and $P_{3f}$ from the coupled ERSJ model are shown in solid lines, and $P_{3f}$ from the uncoupled ERSJ in a dashed line. Inset is the schematic of the probe-sample arrangement.}
\end{figure}

\clearpage

\end{document}